\documentclass[10pt, conference, compsocconf]{IEEEtran}

\usepackage{cite}
\usepackage{graphicx}
\usepackage{amssymb,amsthm}
\usepackage[cmex10]{amsmath}
\usepackage[boxed]{algorithm}
\usepackage{algorithmic}
\usepackage{array}
\usepackage{stfloats}
\usepackage{url}
\usepackage{lipsum}
\usepackage{subfigure}
\usepackage{tabularx}

\makeatletter
\def\thm@space@setup{\thm@preskip=2pt
\thm@postskip=2pt \itshape}
\makeatother
\newtheoremstyle{newstyle}      
{} 
{} 
{\mdseries} 
{} 
{\bfseries} 
{.} 
{ } 
{} 

\theoremstyle{newstyle}

\theoremstyle{definition}

\theoremstyle{remark}

\newcommand{\set}[1]{{\mathcal{{#1}}}}
\newcommand{\abs}[1]{{\left|{#1}\right|}}
\newcommand{\xor}{{\oplus}}
\newcommand{\pc}[1]{{\left\{{#1}\right\}}}
\newcommand{\pr}[1]{{\left({#1}\right)}}

\begin{document}
\sloppy

\title{Coded TeraSort}




%
\author{\IEEEauthorblockN{Songze Li$^{*}$, Sucha Supittayapornpong$^{*}$, Mohammad Ali Maddah-Ali$^{\dagger}$, and Salman Avestimehr$^{*}$}
\IEEEauthorblockA{$^{*}$University of Southern California, $^{\dagger}$Nokia Bell Labs\\
Email: \{songzeli,supittay\}@usc.edu,
mohammadali.maddah-ali@alcatel-lucent.com, avestimehr@ee.usc.edu}}

\maketitle

\begin{abstract}
We focus on sorting, which is the building block of many machine learning algorithms, and propose a novel distributed sorting algorithm, named \texttt{CodedTeraSort}, which substantially improves the execution time of the \texttt{TeraSort} benchmark in Hadoop MapReduce. The key idea of \texttt{CodedTeraSort} is to impose \emph{structured} redundancy in data, in order to enable in-network coding opportunities that overcome the data shuffling bottleneck of \texttt{TeraSort}. We empirically evaluate the performance of \texttt{CodedTeraSort} algorithm on Amazon EC2 clusters, and demonstrate that it achieves $\boldsymbol{1.97\times}$ - $\boldsymbol{3.39\times}$ speedup, compared with \texttt{TeraSort}, for typical settings of interest.

\end{abstract}


\section{Introduction}
\label{sec:intro}
The modern paradigm for large-scale machine learning, scientific computing, and data analysis involves a massively large distributed system comprising individually small and relatively unreliable computing nodes made of commodity low-end hardware.  Specifically, distributed systems like Apache Spark~\cite{zaharia2010spark} and computational primitives like MapReduce~\cite{dean2004mapreduce}, Dryad~\cite{Dryad}, and CIEL~\cite{Ciel} have gained significant traction, as they enable the execution of production-scale tasks on data sizes of the order of tens of terabytes and more.

However, as we ``scale out'' computations across many distributed nodes, massive amounts of raw and (partially) computed data must be moved among nodes, often over many iterations of a running algorithm, to execute the computational tasks. This creates a substantial \emph{communication bottleneck}. For example, by analyzing Hadoop traces from Facebook, it is demonstrated that, on average, 33\% of the overall job execution time is spent on data shuffling~\cite{chowdhury2011managing}. This ratio can be much worse for sorting and other basis tasks underlying many machine learning applications. For example, as shown in~\cite{zhang2013performance}, $50\%$ - $70\%$ of the execution time can be spent for data shuffling in applications including TeraSort, WordCount, RankedInvertedIndex, and SelfJoin.

Recently, it has been been shown that ``coding'' can provide novel opportunities to improve the run-time performance of machine learning applications. On one hand, coding can significantly slash the communication load of distributed computing by leveraging carefully designed redundant local computations at the nodes. In particular, a coding framework for MapReduce, named \emph{Coded MapReduce}, has been proposed in~\cite{LMAall,LMAISIT16,li2016fundamental,li2016unified}, which assigns the
computation of each Map task at $r$ \emph{carefully chosen} nodes (for some $r \in \mathbb{N}$), in order to enable in-network coding
opportunities  that reduce the communication load by  $r\times$. For example,  by redundantly computing each Map task at only \emph{two}  carefully chosen nodes, Coded MapReduce can reduce the communication load of MapReduce by 50\%. On the other hand, it was shown in~\cite{Lee2015Bcast} that error correcting codes (e.g.,  Maximum-Distance-Separable codes) can be utilized to create redundant computation tasks for linear computations (e.g., matrix multiplication), which can effectively alleviate the straggler effect in many distributed machine learning algorithms. For example, as demonstrated in~\cite{Lee2015Bcast}, coded computing reduces the average run-time of a gradient descent algorithm for linear regression by 31.3\% to 35.7\%.

The goal of this paper is to \emph{demonstrate as a proof of concept the impact of coding in reducing the data shuffling load of distributed computing, and speeding up the overall computations}. We focus on ``sorting'', which is not only a basic benchmark for distributed computing systems like Hadoop MapReduce and Spark, but also a key step in many machine learning algorithms including recommender systems, SVD and many graph algorithms, and has data shuffling as its main bottleneck. Our main result is the development of a new distributed sorting algorithm, named \texttt{CodedTeraSort}, that imposes \emph{structured} redundancy in data to enable coding opportunities for efficient data shuffling, which results in speeding up the state-of-the-art algorithms by $1.97\times$- $3.39\times$ in typical settings of interest.

To date, there have been many distributed sorting algorithms developed to perform efficient distributed sorting on commodity hardware (see, e.g.,~\cite{akl2014parallel,pasetto2011comparative}). Out of these algorithms, \texttt{TeraSort}~\cite{TSpackage}, originally developed to sort terabytes of data~\cite{Terasort}, is a commonly used benchmark in Hadoop MapReduce~\cite{had}.  In consistence with the general structure of a MapReduce execution, in a \texttt{TeraSort} execution, each server node first \emph{maps} each data point it stores locally into a particular partition of the key space, then all the data points in the same partition are \emph{shuffled} to a single node, on which they are sorted within the partition to \emph{reduce} the final sorted output. 

Out of the above three steps, the time spent in the Map and the Reduce phases of the computation can be reduced by paralleling onto more processing nodes, while the shuffle time will almost remain constant. This is because that no matter how large the cluster size is, almost as much as the entire raw dataset of data need to be transferred over the network. Hence, data shuffling often becomes the bottleneck of the performance of the \texttt{TeraSort} algorithm (see, e.g.,~\cite{guo2013ishuffle,zhang2013performance}).   

In this paper, we propose to leverage coding to overcome the shuffling bottleneck of \texttt{TeraSort}. In particular, we develop a novel distributed sorting algorithm, named \texttt{CodedTeraSort}, that incorporates the coding ideas in~\cite{li2016fundamental} to inject structured computation redundancy in Map phase of \texttt{TeraSort}, in order to cut down its shuffling load.  At a high-level \texttt{CodedTeraSort} can be explained as the following: 
\begin{itemize}
    \item The input data points are split into disjoint  files, and each file is stored on \emph{multiple carefully selected} server nodes to create \emph{structured redundancy} in data.
    \item Each node maps all files that are assigned to it, following the Map procedure of \texttt{TeraSort}.
    \item Each node utilizes the imposed structured redundancy in data placement to create \emph{coded packets} for data shuffling, such that the \emph{multicast} of each coded packet delivers data points to several nodes simultaneously, hence speeding up the data shuffling phase. 
    \item Each node decodes the data points that it needs for Reduce phase, from the received coded packets, and follows the Reduce procedure of \texttt{TeraSort}.
\end{itemize}


We empirically evaluate the performance of \texttt{CodedTeraSort} through extensive experiments over Amazon EC2 clusters. While the underlying EC2 networking does not support network-layer multicast, we perform the application-layer multicast for shuffling of coded packets, using the broadcast API \texttt{MPI\_Bcast} from Open~MPI~\cite{openMPI}. Compared with the conventional \texttt{TeraSort} implementation, we demonstrate that \texttt{CodedTeraSort} achieves  $1.97\times$-$3.39\times$ speedup for typical settings of interest. Despite the extra overhead imposed by coding (e.g., generation of the coding plan, data encoding and decoding) and application-layer multicasting, the practically achieved performance gain approximately matches the gain theoretically promised by the \texttt{CodedTeraSort} algorithm.   



\section{Overview of Coded MapReduce (CMR)}
\label{sec:overview}
In this section, we provide an overview of the recently proposed \emph{Coded MapReduce} (CMR) framework~\cite{LMAall,LMAISIT16,li2016fundamental} via an example, to illustrate how coding can be utilized to reduce the data shuffling load in distributed computing.

Let us consider a general MapReduce-type framework for distributed computing, in which the overall computation is decomposed to three stages, \emph{Map}, \emph{Shuffle}, and \emph{Reduce} that are executed distributively across many computing nodes. In the Map stage, each input file is processed locally, in one (or more) of the nodes, to generate \emph{intermediate values}. In the Shuffle stage, for every output function to be calculated, all intermediate values corresponding to that function are transferred to one of the nodes for reduction. Finally, in the Reduce stage all intermediate values of a function are reduced to the final result.

The driving idea for CMR is to leverage the available or under-utilized computing  resources at various parts of the network, in order to create structured redundancy in computations that provides in-network coding opportunities for significantly reducing the data shuffling load. Let us illustrate CMR via a simple example.


\begin{figure}[t]
  \centering
  \subfigure[Uncoded MapReduce Scheme.]{\includegraphics[width=0.35\textwidth, trim = 0cm 0cm 0cm 0cm]{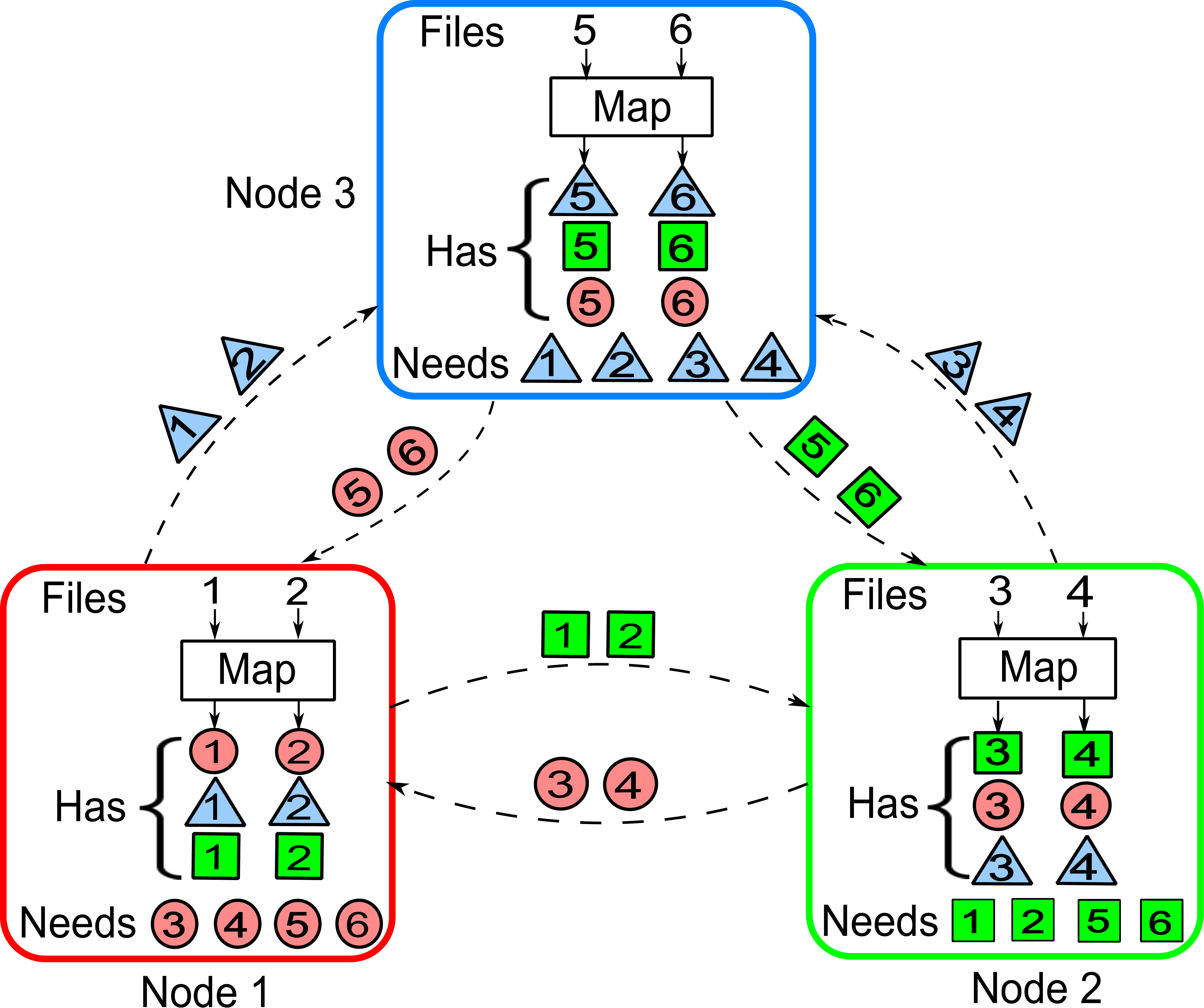}
      \label{fig:UncodedMR}}
      \subfigure[Coded MapReduce Scheme.]{\includegraphics[width=0.4\textwidth]{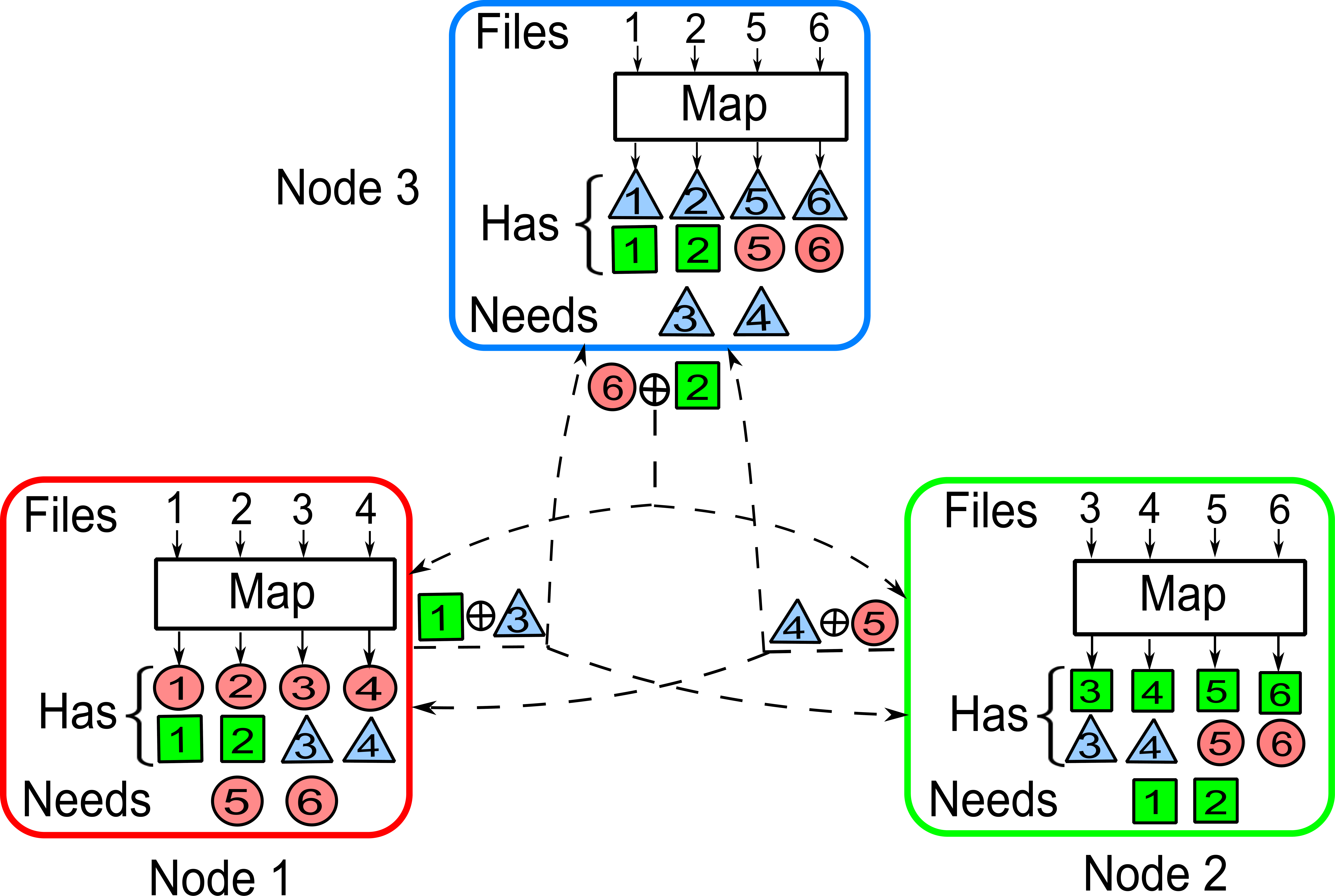}
      \label{fig:CMR}}
  \caption{ Illustrations of the conventional uncoded MapReduce and the Coded MapReduce schemes to compute $Q=3$ functions from $N=6$ inputs using $K=3$ nodes.}
  \label{fig:MR}
\end{figure}

\noindent \textbf{Example (Coded MapReduce).}
Consider the MapReduce problem in Fig.~\ref{fig:MR} for distributed computing of $3$  output functions, represented by red/circle, green/square, and blue/triangle respectively, from $6$ input files, using three computing nodes. Nodes~$1$, $2$, and $3$ are respectively responsible for final reduction of red/circle, green/square, and  blue/triangle output functions. Let us first consider the case where no redundancy is imposed on the computations, i.e., each file is mapped once. As shown in Fig.~\ref{fig:UncodedMR}, Node $i$ maps files $2i-1$ and $2i$ for $i=1,2,3$. In this case, each node maps $2$ input files locally\footnote{Note that when a node maps a file, it computes all three intermediate values of that file needed for the three output functions.}, obtaining $2$ out of $6$ required intermediate values to reduce its output function. Hence, each node needs $4$ intermediate values  from the other nodes, yielding a communication load of $4 \times 3=12$.

Now, we demonstrate how to leverage computation redundancy to slash the communication load via in-network coding. As shown in Fig.~\ref{fig:CMR}, computation load is doubled such that each file is now mapped on two nodes (files are cached prior to computations at the nodes). It is apparent that since more local computations are performed, each node now only requires $2$ other intermediate values, and an uncoded shuffling scheme would achieve a communication load of $2 \times 3=6$. However, we can do much better with coding. As shown in Fig.~\ref{fig:CMR}, instead of unicasting individual intermediate values, every node multicasts XOR, denoted by $\oplus$, of $2$ intermediate values to the other two nodes, simultaneously satisfying their data demands\footnote{This type of coding was also utilized to solve the index coding problem~\cite{bar2011index,el2010index} that arises from the network coding problem~\cite{ahlswede2000network}.}.  For example, knowing the blue/triangle in File~$3$, Node~$2$ can cancel it from the coded packet sent by Node~$1$, recovering the needed green/square in File~$1$. Therefore, this coding incurs a communication load of $3$, achieving a $2\times$ gain from the uncoded shuffling.
$\hfill \square$



More generally, we can consider a distributed computing scenario, where $K$ nodes collaborate to  compute  $Q$ arbitrary functions ($\phi_1,\ldots,\phi_Q$) from $N$ inputs ($x_1,\ldots,x_N$) via a MapReduce-type framework, for $q=1,\ldots,Q$:
\begin{equation}
\phi_q(x_1,\ldots,x_N) = \underbrace{g_q}_\text{reduce}(\underbrace{\ell_{q,1}}_\text{map}(x_1),\ldots,\underbrace{\ell_{q,N}}_\text{map}(x_N)).    
\end{equation}
For this scenario, the \emph{computation load}, $r$, is defined as the average number of nodes that map each input file (e.g., $r = 2$ in the example of Fig.~\ref{fig:CMR} since each file is mapped on two nodes). Similarly, the \emph{communication load}, $L$, is defined as the total amount of intermediate values (i.e., $\ell_{q,j}(x_j)'s$) that need to be exchanged across nodes in the data shuffling stage (normalized by $QN$), in order to compute all $Q$ output functions. 

\begin{figure}[htbp]
  \centering
  \includegraphics[width=0.4\textwidth]{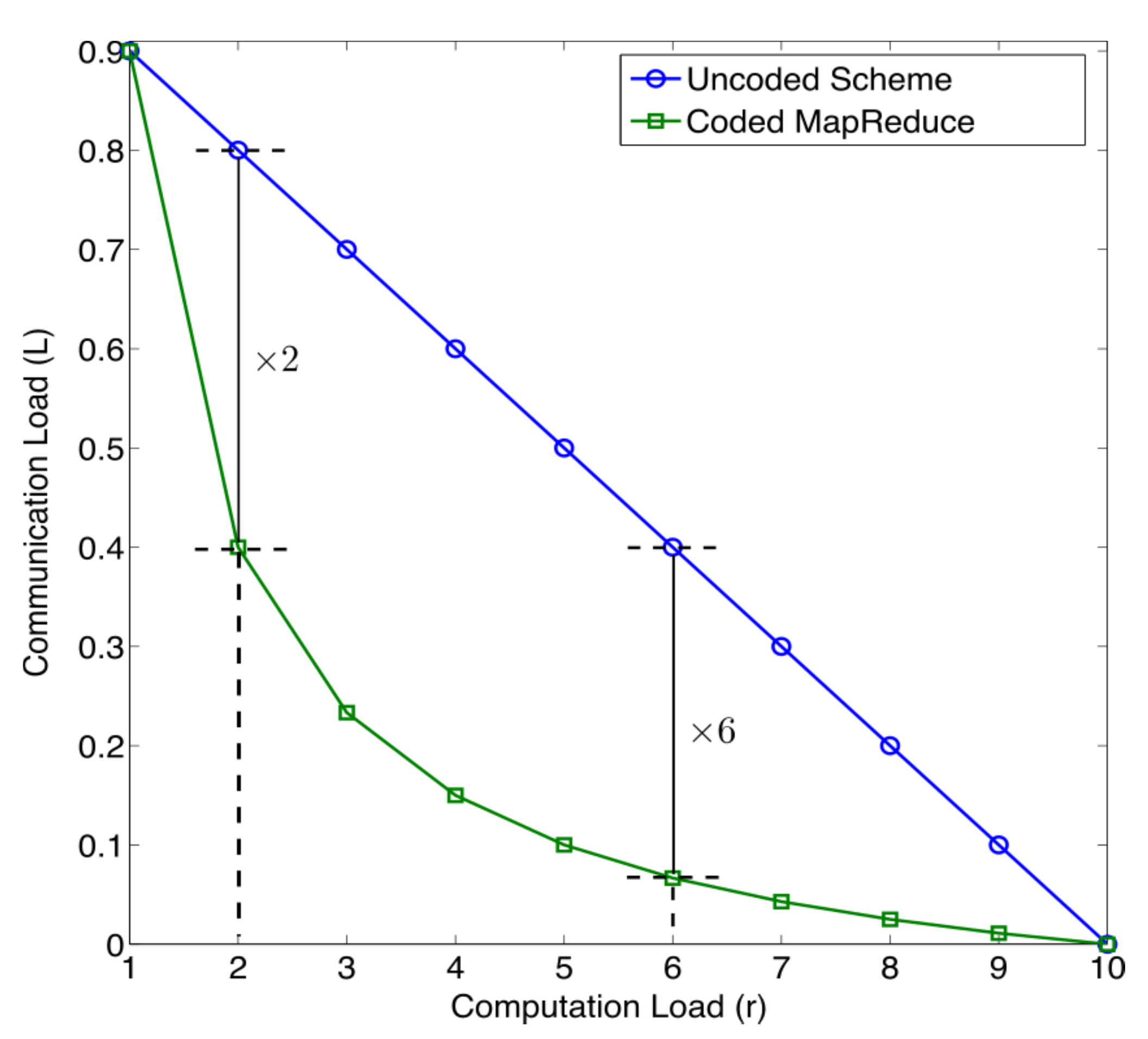}
  \caption{Comparison of the communication load of CMR with that of the uncoded schemes. For integer-valued computation load $r$, CMR achieves $r$ times smaller communication load. Figure from~\cite{li2016fundamental}. }
  \label{fig:scaling}
\end{figure}

In the general CMR scheme proposed in~\cite{LMAall,li2016fundamental}, the computation of each Map task is repeated at $r$ \emph{carefully chosen} nodes (i.e., incurring computation load of $r$), in order to enable the nodes to exchange coded multicast messages that are \emph{simultaneously} useful for $r$ other nodes. As a result, CMR reduces the communication load by exactly a multiplicative factor of the computation load $r$ (see Fig.~\ref{fig:scaling}), i.e., achieving
\begin{equation}
L_{\textup{CMR}}(r)= \tfrac{1}{r}L_{\textup{uncoded}}(r)=\tfrac{1}{r}(1-\tfrac{r}{K})= \Theta(\tfrac{1}{r}).\label{eq:tradeoff}    
\end{equation}

Also in~\cite{li2016fundamental}, an information-theoretic lower bound on the minimum possible communication load, $L^*(r)$, was derived, which was demonstrated to exactly match that achieved by CMR, i.e., $L^*(r) = L_{\textup{CMR}}(r)=\Theta(\tfrac{1}{r})$. Interestingly, this has revealed a fundamental \emph{inversely-linear proportional} tradeoff between computation load ($r$) and communication load ($L$),  which can be utilized to optimally trade the available or under-utilized computing resources in the network for communication bandwidth. 

CMR can also reduce the overall execution time of MapReduce by balancing the computation load in the Map stage and the communication load in the Shuffle stage. To illustrate this, let us consider a MapReduce application for which the overall response time is composed of the time spent executing the Map tasks, denoted by $T_{\textup{map}}$, the time spent shuffling intermediate values, denoted by $T_{\textup{shuffle}}$, and the time spent executing the Reduce tasks, denoted by $T_{\textup{reduce}}$, i.e.,
\begin{equation}
\label{eq:totalMR}
T_{\textup{total, MR}} = T_{\textup{map}}+ T_{\textup{shuffle}}+ T_{\textup{reduce}}.
\end{equation}
Using CMR, we can leverage $r\times$ more computations in the Map phase, in order to reduce the communication load by the same multiplicative factor, where $r \in \mathbb{N}$ is a design parameter that can be optimized to minimize the overall execution time. Hence, CMR promises that we can achieve the overall execution time of
\begin{align}\label{eq:total}
T_{\textup{total, CMR}} \approx rT_{\textup{map}}+ \tfrac{1}{r}T_{\textup{shuffle}}+ T_{\textup{reduce}},
\end{align}
for any $1 \leq r \leq K$, where $K$ is the total number of nodes on which the distributed computation is executed. To minimize the above execution time, one would choose $$r^* =\left\lfloor \sqrt{\tfrac{T_{\textup{shuffle}}}{T_{\textup{map}}}} \right\rfloor \textup{ or } \left\lceil \sqrt{\tfrac{T_{\textup{shuffle}}}{T_{\textup{map}}}} \right\rceil, $$  resulting in execution time of
\begin{equation}
\label{eq:totalCMR}
T^*_{\textup{total, CMR}} \approx 2\sqrt{T_{\textup{shuffle}}T_{\textup{map}}}+T_{\textup{reduce}}.
\end{equation}
For example, in a MapReduce application that $T_{\textup{shuffle}}$ is $10\times$ - $100\times$ larger than $T_{\textup{map}}+T_{\textup{reduce}}$,  by comparing from (\ref{eq:totalMR}) and (\ref{eq:totalCMR}), we note that CMR can reduce the execution time by approximately $1.5 \times$ - $5 \times$. 

In the rest of this paper, we demonstrate how to utilize the ideas from CMR in order to develop a new distributed sorting algorithm, named \texttt{CodedTeraSort}, that leverages coding to speedup the conventional sorting algorithm, \texttt{TeraSort}. We will also empirically demonstrate the performance of \texttt{CodedTeraSort} via experiments over Amazon EC2 clusters.

\section{TeraSort}
\label{sec:TeraSort}
\texttt{TeraSort}~\cite{Terasort} is a conventional algorithm for distributed sorting of a large amount of data. The input data that is to be sorted is in the format of key-value (KV) pairs, meaning each input KV pair consists of a key and a value. For example, the domain of the keys can be 10-byte integers, and the domain of the values can be arbitrary strings. \texttt{TeraSort} aims to sort the input data according to their keys, e.g., sorting integers. 

\begin{figure}
  \centering
  \includegraphics[scale=0.83]{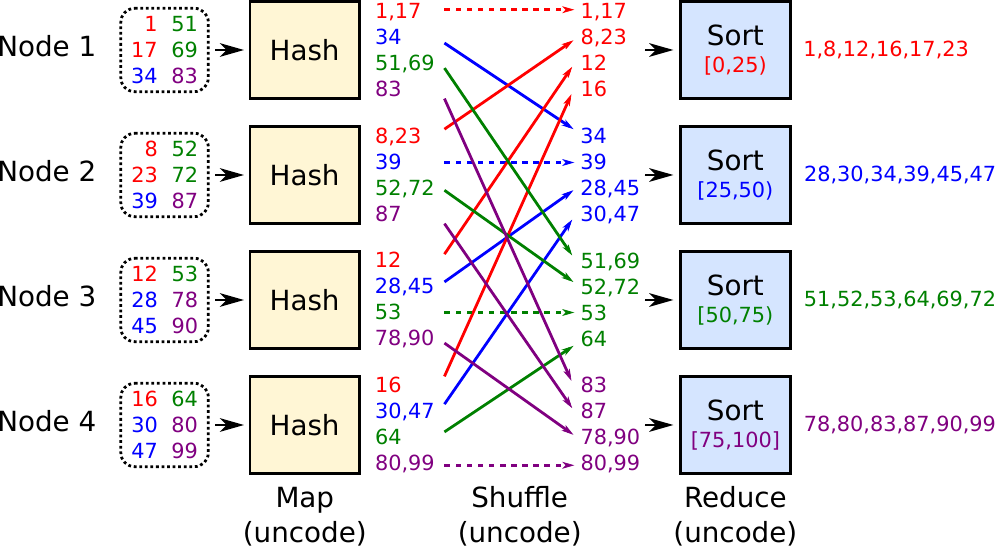}
  \caption{Illustration of conventional \texttt{TeraSort} with $K = 4$ nodes and key domain partitions $[0,25), [25,50), [50,75), [75,100]$.  A dotted box represents an input file.  An input file is hashed into 4 groups of KV pairs, one for each partition. For each of the 4 partitions, the 4 groups of KV pairs belonging to that partition computed on 4 nodes are all fetched to a corresponding node, which sorts all KV pairs in that partition locally.}
  \label{fig:uncode}
\end{figure}

Let us consider \texttt{TeraSort} for distributed sorting over $K$ nodes, whose indices are denoted by a set ${\cal K} =\{1,\ldots,K\}$. The implementation consists of 5 components: File Placement, Key Domain Partitioning, Map Stage, Shuffle Stage, and Reduce Stage.  In File Placement, the entire KV pairs are split into $K$ disjoint files, and each file is placed on one of the $K$ nodes.  In Key Domain Partitioning, the domain of the keys is split into $K$ partitions, and each node will be responsible for sorting the KV pairs whose keys fall into one of the partitions. In Map Stage, each node hashes each KV pair in its locally stored file into one of the $K$ partitions, according to its key. In Shuffle Stage, the KV pairs in the same partition are delivered to the node that is responsible for sorting that partition. In Reduce Stage, each node locally sorts KV pairs belonging to its assigned partition.  A simple example illustrating \texttt{TeraSort} is shown in Fig. \ref{fig:uncode}.  We next discuss each component in detail.

\subsection{Algorithm Description}

\subsubsection{File Placement}
Let $F$ denote the entire KV pairs to be sorted.  They are split into $K$ disjoint input files, denoted by $F_\pc{1}, \dotsc, F_\pc{K}$.  
File $F_\pc{k}$ is assigned to and locally stored at Node $k$.

\subsubsection{Key Domain Partitioning}
The key domain of the KV pair, denoted by $P$, is split into $K$ \emph{ordered} partitions, denoted by $P_1, \dotsc, P_K$. Specifically, for any $p \in P_i$ and any $p' \in P_{i+1}$, it holds that $p < p'$ for all $i \in \pc{1, \dotsc, K-1}$.  For example, when $P = [0, 100]$ and $K = 4$, the partitions can be $P_1 = [0, 25), P_2 = [25, 50), P_3 = [50, 75), P_4 = [75, 100]$. Node $k$ is responsible for sorting all KV pairs in the partition $P_k$, for all $k \in \set{K}$.

\subsubsection{Map Stage}\label{sec:map}
In this stage, each node hashes each KV pair in the locally stored file $F_\pc{k}$ to the partition its key falls into.  For each of the $K$ key partitions, the hashing procedure on the file $F_\pc{k}$ generates an \emph{intermediate value} that contains the KV pairs in $F_\pc{k}$ whose keys belong to that partition.
%
%
More specifically, we denote the intermediate value of the partition $P_j$ from the file $F_\pc{k}$ as $I_\pc{k}^j$, and the hashing procedure on the file $F_\pc{k}$ is defined as 
\begin{equation*}
    \pc{ I_\pc{k}^1, \dotsc, I_\pc{k}^K } \leftarrow Hash\pr{ F_\pc{k} }.
\end{equation*}

\subsubsection{Shuffle Stage}
During this stage, 
the intermediate value $ I_\pc{j}^k$ calculated at Node $j$, $j \neq k$, is unicast to Node~$k$ from Node $j$, for all $k \in \set{K}$. Since the intermediate value $I_\pc{k}^k$ is computed locally at Node $k$ in the Map stage, by the end of the Shuffle stage, Node $k$ knows all intermediate values ${\pc{I_\pc{1}^k, \dotsc, I_\pc{K}^k}}$ of the partition $P_k$ from all $K$ files.

\subsubsection{Reduce Stage}\label{sec:reduce}
In this stage, Node $k$ locally sorts all KV pairs whose keys fall into the partition $P_k$, for all $k\in \set{K}$.  Specifically, it sorts all intermediate values in the partition $P_k$ into a sorted list $Q_k$ as follows
\begin{equation*}
    Q_k \leftarrow Sort\pr{ {\pc{I_\pc{1}^k, \dotsc, I_\pc{K}^k} } }.
\end{equation*}

Since the partitions are created in the ascending order as specified in the above Key Domain Partitioning step, the collection of the $K$ sorted list generated in the Reduce stage, i.e., $(Q_1,\ldots,Q_K)$ represents the final sorted list of the entire input data.


\subsection{Performance Evaluation}\label{sec:TeraSortPerf}
To understand the performance of \texttt{TeraSort}, we performed an experiment on Amazon EC2 to sort 12GB of data by running \texttt{TeraSort} on 16 nodes. The breakdown of the total execution time is shown in Table \ref{tlb:exampleTeraSort}.

\begin{table}[t!]
    \centering  
    \caption{Performance of \texttt{TeraSort} sorting $12$GB data with $K=16$ nodes and $100$ Mbps network speed} 
    \label{tlb:exampleTeraSort}
    \begin{tabular}{|c|c|c|c|c|c|}\hline
    Map    & Pack   & Shuffle  & Unpack  & Reduce  & Total \\
    (sec.) & (sec.) & (sec.)   & (sec.)  & (sec.)  & (sec.) \\\hline
    1.86   & 2.35   & 945.72   & 0.85    & 10.47   & 961.25 \\\hline
    \end{tabular}
\end{table}

We observe from Table \ref{tlb:exampleTeraSort} that for a conventional \texttt{TeraSort} execution, 98.4\% of the total execution time was spent in data shuffling, which is $508.5 \times$ of the time spent in the Map stage. Given the fact that data shuffling dominates the job execution time, the principle of optimally trading computation for communication of Coded MapReduce reviewed in Section~\ref{sec:overview} can be applied to significantly improve the performance of \texttt{TeraSort}. Following the theoretical characterization of the total execution time achieved by CMR in (\ref{eq:total}), when executing the same sorting job using a coded version of \texttt{TeraSort} with a computation load of $r^* = \left\lceil \sqrt{\tfrac{T_{\textup{shuffle}}}{T_{\textup{map}}}} \right\rceil= 23$ (i.e., each input file is repeatedly mapped on $23$ servers), we could theoretically save the total execution time by approximately $10 \times$. This great promise of using CMR to improve the performance of \texttt{TeraSort} motivates us to develop a novel coded distributed sorting algorithm, named \texttt{CodedTeraSort}, which integrates the coding technique of CMR into the above described \texttt{TeraSort} algorithm to reduce the total execution time. We describe \texttt{CodedTeraSort} in detail in the next section.

\section{Coded TeraSort}
\label{sec:codedTeraSort}
In this section, we describe the \texttt{CodedTeraSort} algorithm, which is developed by integrating the coding techniques of the Coded MapReduce scheme illustrated in Section~\ref{sec:overview} into the above described \texttt{TeraSort} algorithm. \texttt{CodedTeraSort} exploits redundant computations on the input files in the Map stage, enabling in-network coding opportunities to significantly slash the load of data shuffling.

\texttt{CodedTeraSort} sorts a group of input KV pairs distributedly over $K$ nodes, through the following $6$ stages of operations:
\begin{enumerate}
    \item \emph{Structured Redundant File Placement}. The entire input KV pairs are split into many small files, each of which is repeatedly placed on multiple nodes according to a particular pattern. 
    \item \emph{Map}. Each node applies the hashing operation as in \texttt{TeraSort} on each of its assigned files.
    \item \emph{Encoding to Create Coded Packets}. Each node generates coded multicast packets from local results computed in Map stage.
    \item \emph{Multicast Shuffling}. Each node multicasts each of its generated coded packet to a specific set of other nodes.
    \item \emph{Decoding}. Each node locally decodes the required KV pairs from the received coded packets.
    \item \emph{Reduce}. Each node locally sorts the KV pairs within its assigned partition.
\end{enumerate}

Next, we  describe the above $6$ stages in detail.





\subsection{Structured Redundant File Placement}
\label{sec:codefile}
For some parameter $r \in \pc{1, \dotsc, K}$, we first split the entire input KV pairs into $N = \binom{K}{r}$ input files. Unlike the file placement of \texttt{TeraSort}, \texttt{CodedTeraSort} places each of the $N$ input files \emph{repetitively} on $r$ distinct nodes.   

We label an input file using a unique subset $\set{S}$ of ${\cal K}$ with size $|\set{S}|=r$, i.e., the $N$ input files are denoted by
\begin{align}
\{F_{\cal S}: {\cal S} \subseteq {\cal K}, |{\cal S}|=r \}.
\end{align}
For example, when $K=4$ and $r=2$, the set of the input files is $\big\{F_{\{1,2\}}, F_{\{1,3\}}, F_{\{1,4\}}, F_{\{2,3\}}, F_{\{2,4\}}, F_{\{3,4\}}\big\}$. 

\begin{figure}[t!]
    \centering
    \includegraphics[scale=1.2]{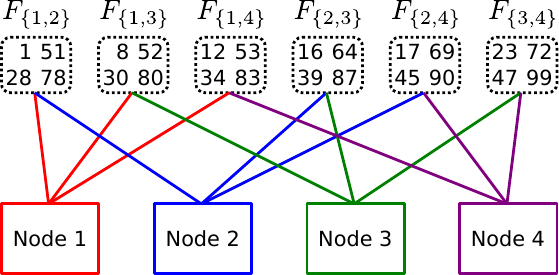}
    \caption{An illustration of the structured redundant file placement in \texttt{CodedTeraSort} with $K = 4$ nodes and $r = 2$.}
    \label{fig:codeplacement}
\end{figure}

We repetitively place an input file $F_{\cal S}$ on each of the $r$ nodes in ${\cal S}$, and hence each node now stores $Nr/K = \binom{K-1}{r-1}$ files.  As illustrated in a simple example in Fig.~\ref{fig:codeplacement} for $K=4$ and $r=2$, the file $F_\pc{2,3}$ is placed on Nodes $2$ and $3$. Node $2$ has files $F_\pc{1,2}, F_\pc{2,3}, F_\pc{2,4}$. We note that this redundant file placement strategy induces a \emph{structured} distribution of the input files such that every subset of $r$ nodes have a unique file in common.

As is done in the \texttt{TeraSort}, the key domain of the input KV pairs is split into $K$ ordered partitions $P_1,\ldots,P_K$, and Node $k$ is responsible for sorting all KV pairs in the partition $P_k$ in the Reduce stage, for all $k \in \set{K}$.

\subsection{Map}\label{sec:codemap}
In this stage, each node repeatedly performs the Map stage operation of \texttt{TeraSort} described in Section \ref{sec:map}, on each input file placed on that node.  Specifically, for each file $F_\set{S}$ with $k \in \set{S}$ that is placed on Node $k$, Node $k$ hashes the KV pairs in $F_\set{S}$ to generate a set of $K$ intermediate values $\pc{I_{\cal S}^1, \dotsc, I_{\cal S}^K}$.

\begin{figure}
    \centering
    \includegraphics[scale=1.0]{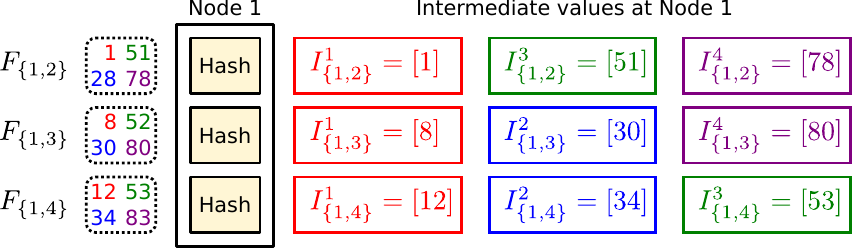}
    \caption{An illustration of the Map stage at Node 1 in \texttt{CodedTeraSort} with $K = 4$, $r = 2$ and the key partitions $[0,25), [25,50), [50,75), [75,100]$.}
    \label{fig:codemap}
\end{figure}

Only relevant intermediate values generated in the Map stage are kept locally for further processing.  In particular, out of the $K$ intermediate values $\pc{I_{\cal S}^1, \dotsc, I_{\cal S}^K}$ generated from file $F_{\cal S}$, only $I_{\cal S}^k$ and $\pc{I_{\cal S}^i : i \in \set{K} \backslash {\cal S}}$ are kept at Node $k$.  This is because that the intermediate value $I_{\cal S}^i$, required by Node~$i \in {\cal S} \backslash \{k\}$ in the Reduce stage, is already available at Node $i$ after the Map stage, so Node $k$ does not need to keep them 
and send them to the nodes in ${\cal S} \backslash \{k\}$.  For example, as shown in Fig. \ref{fig:codemap}, Node 1 does not keep the intermediate value $I_\pc{1,2}^2$ for Node 2. However, Node $1$ keeps $I_\pc{1,2}^1, I_\pc{1,2}^3, I_\pc{1,2}^4$, which are required by Nodes 1, 3, and 4 in the Reduce stage. 


\subsection{Encoding to Create Coded Packets}
After the Map stage, each node has known locally a part of the KV pairs in the partition it is responsible for sorting, i.e., $\{I_{\cal S}^k: k \in {\cal S}, |\set{S}|=r\}$ for Node $k$. In the next stages of the computation, the server nodes need to communicate with each other to exchange the rest of the required intermediate values to perform local sorting in the Reduce stage. 

The role of the encoding process is to exploit the structured data redundancy created by the particular repetitive file placement described above, in order to create coded multicast packets that are simultaneously useful for multiple nodes, thus saving the load of communicating intermediate values. For example in Fig.~\ref{fig:codemap}, Node 1 wants to send $I^2_{\{1,3\}}=[30]$ to Node 2 and $I^3_{\{1,2\}}=[51]$ to Node 3. Since $I^3_{\{1,2\}}$ and $I^2_{\{1,3\}}$ are already known at Node 2 and Node 3 respectively after the Map stage, instead of unicasting these two intermediate values individually, Node 1 can rather multicast a coded packet generated by XORing these two values, i.e., $[30\oplus 51]$. Then Node 2 and 3 can decode their required intermediate values using locally known intermediate values, e.g., Node 2 uses $I_\pc{1,2}^3 = [51]$ to decode $I^2_{\{1,3\}}$ by computing $I^2_{\{1,3\}}=[30 \oplus 51] \oplus [51] = [30]$. By multicasting a coded packet instead of unicasting two uncoded ones, we save the load of communication by 50\%. 

More generally, in the encoding stage, every node creates coded packets that are simultaneously useful for $r$ other nodes. Specifically, in every subset $\set{M} \subseteq \set{K}$ of $|\set{M}|=r+1$ nodes, the encoding operation proceeds as follows.
\begin{itemize}
    \item For each $t \in \set{M}$, the intermediate value $I^t_{\set{M} \backslash \{t\}}$, which is know at all nodes in $\set{M} \backslash \{t\}$, is evenly and arbitrarily split into $r$ segments, i.e., 
    \begin{align}
    I^t_{\set{M} \backslash \{t\}} = \{ I^t_{\set{M} \backslash \{t\},k}: k\in \set{M} \backslash \{t\} \},
    \end{align}
    where $I^t_{\set{M} \backslash \{t\},k}$ denotes the segment corresponding to Node $k$.
    \item For each $k \in \set{M}$, we generate the coded packet of Node $k$ in $\set{M}$, denoted by $E_{\set{M},k}$, by XORing all segments corresponding to Node $k$ in $\set{M}$, \footnote{All segments are zero-padded to the length of the longest one.} i.e.,
    \begin{align}\label{eq:encode}
    E_{\set{M},k} = \underset{{t \in \set{M} \backslash \{k\}}}{\oplus} I^t_{\set{M} \backslash \{t\},k}. 
    \end{align}
\end{itemize}

By the end of the Encoding stage, for each $k \in \set{K}$, Node $k$ has generated $\binom{K-1}{r}$ coded packets, i.e., $\{E_{{\cal M},k}: k \in \set{M}, |\set{M}|=r+1\}$. 

\begin{figure}
    \centering
    \includegraphics[scale=0.8]{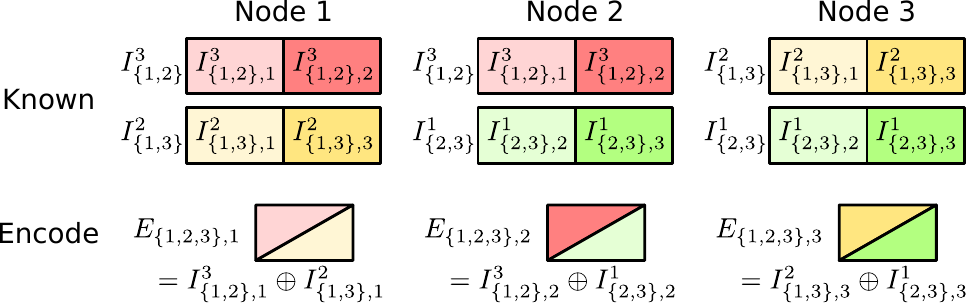}
    \caption{An illustration of the encoding process within a multicast group ${\cal M}=\pc{1,2,3}$.}
    \label{fig:codeencode}
\end{figure}

In Fig.~\ref{fig:codeencode}, we consider a scenario with $r=2$, and illustrate the encoding process in the subset ${\cal M}=\{1,2,3\}$. Exploiting the particular structure imposed in the stage of file placement, each node creates a coded packet that contains data segments useful for the other 2 nodes.

We summarize the the pseudocode of the Encoding stage at Node $k$ in Algorithm \ref{alg:encode}.

\begin{algorithm}
  \begin{algorithmic}
    \STATE{\texttt{// At Node $k$ //}}\vspace{1mm}
    \STATE{\texttt{// Data Segmentation}}
    \FOR{each $\set{M} \subseteq \set{K}$ with $|\set{M}|=r+1$ and $k \in \set{M}$}
      \FOR{each $t \in \set{M} \backslash \{k\}$}
      \STATE{Consider file index $\set{F} \leftarrow \set{M} \backslash\{t\}$}
      \STATE{Evenly split $I_{\set{F}}^t$ to $r$ segments $\{ I_{\set{F},j}^t : j \in \set{F} \}$}
      \ENDFOR
    \ENDFOR\vspace{1mm}
    \STATE{\texttt{// Encode}}
    \FOR{each $\set{M} \subseteq \set{K}$ with $|\set{M}|=r+1$ and $k \in \set{M}$}
    \STATE{Initialize coded packet $E_{\set{M},k} \leftarrow \emptyset$}
      \FOR{each $t \in \set{M} \backslash \{k\}$}
      \STATE{Consider file index $\set{F} \leftarrow \set{M}\backslash\{t\}$}
      \STATE{$E_{\set{M},k} \leftarrow E_{\set{M},k} \xor I_{\set{F},k}^t$}
      \STATE{Store $\pc{ I_{\set{F},j}^t : j \in \set{F}\backslash\pc{k}}$}
      \ENDFOR
    \STATE{Store $E_{\set{M},k}$}
    \ENDFOR
  \end{algorithmic}
  \caption{Encoding to Create Coded Packets}
  \label{alg:encode}
\end{algorithm}

\subsection{Multicast Shuffling}
After all coded packets are created at the $K$ nodes, the Multicast Shuffling process takes place within each subset of $r+1$ nodes. Specifically, within each group $\set{M} \subseteq \set{K}$ of $|\set{M}|=r+1$ nodes, each Node $k \in \set{M}$ multicasts its coded packet $E_{{\cal M},k}$ to the other nodes in $\set{M} \backslash \{k\}$.  

As we have seen in the encoding process, each coded packet is simultaneously useful for $r$ other nodes. Therefore, compared with an uncoded shuffling scheme that solely uses unicast communications, the multicast shuffling employed by \texttt{CodedTeraSort} reduces the communication load by exactly $r \times$. This gain is even higher compared with the \texttt{TeraSort} algorithm, for which no computation is repeated in the Map stage. This is because that even without multicasting, the redundant computations performed in the Map stage of \texttt{CodedTeraSort} already accumulate more locally available data needed for reduction, requiring less data to be shuffled across the network. 





\begin{figure}
    \centering
    \includegraphics[scale=0.8]{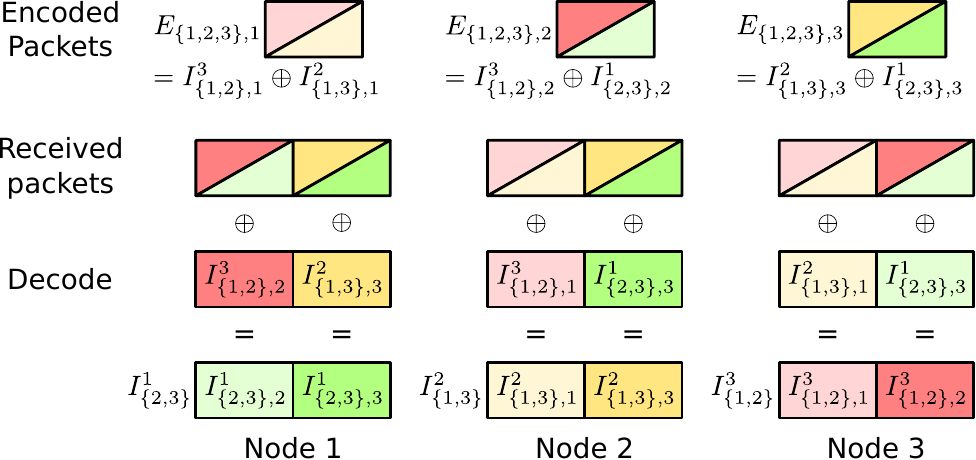}
    \caption{An illustration of the decoding process within a multicast group ${\cal M}=\pc{1,2,3}$.}
    \label{fig:codedecode}
\end{figure}

\subsection{Decoding}
During the stage of Multicast Shuffling, within each multicast group $\set{M} \subseteq \set{K}$ of $|\set{M}|=r+1$ nodes, each Node $k \in \set{M}$ receives a coded packet $E_{\set{M},u}$ from Node $u$, for all $u \in \set{M} \backslash \{k\}$. By the encoding process in (\ref{eq:encode}), we have
  \begin{align}
    E_{\set{M},u} = \underset{{t \in \set{M} \backslash \{u\}}}{\oplus} I^t_{\set{M} \backslash \{t\},u}. 
\end{align}

It is apparent that for all $t \in \set{M} \backslash \{u,k\}$, we have $k \in \set{M} \backslash \{t\}$, and Node $k$ knows locally the intermediate values $I^t_{\set{M} \backslash \{t\}}$, for all $t \in \set{M} \backslash \{u,k\}$, from the Map stage. Therefore, it knows locally all the data segments $\{I^t_{\set{M} \backslash \{t\},u}: t \in \set{M} \backslash \{u,k\}\}$. Then Node $k$ performs the decoding process by XORing these data segments with $ E_{\set{M},u}$, i.e., 
\begin{align}
E_{\set{M},u} \oplus \pr{\underset{{t \in \set{M} \backslash \{u,k\}}}{\oplus} I^t_{\set{M} \backslash \{t\},u}} = I^k_{\set{M} \backslash \{k\},u},
\end{align}
which recovers the data segment $I^k_{\set{M} \backslash \{k\},u}$. 

Similarly, Node $k$ recovers all data segments $\{I^k_{\set{M} \backslash \{k\},u}: u \in \set{M} \backslash \{k\}\}$ from the received coded packets in $\set{M}$, and merge them back to obtain a required intermediate value $I^k_{\set{M} \backslash \{k\}}$.

Finally, we repeat the above decoding process for all subsets of size $r+1$ that contain $k$, and Node $k$ decodes the intermediate values $\{I^k_{\set{M} \backslash \{k\}}: k\in\set{M}, |\set{M}|=r+1\}$, which can be equivalently represented by  
$\pc{ I_\set{S}^k : k \notin \set{S}, \abs{\set{S}} = r }$.


In Fig.~\ref{fig:codedecode}, we consider a scenario with $r=2$, and illustrate the above described decoding process in the subset ${\cal M}=\{1,2,3\}$. In this example, each node receives a multicast coded packet from each of the other two nodes. Each node decodes 2 data segments from the received coded packets, and merge them to recover a required intermediate value.

We summarize the the pseudocode of the Decoding stage at Node $k$ in Algorithm \ref{alg:decode}. 

\begin{algorithm}
  \begin{algorithmic}
    \STATE{\texttt{// At Node $k$ //}}\vspace{1mm}
    \FOR{each $\set{M} \subseteq \set{K}$ with $|\set{M}|=r+1$ and $k \in \set{M}$}
    \STATE{Consider file index $\set{F} \leftarrow \set{M} \backslash\pc{k}$}
      \FOR{each $u \in \set{F}$}
      \STATE{Initialize decoded segment $D_{\set{F},u}^k \leftarrow E_{\set{M},u}$}
        \FOR{each Node $t \in \set{F} \backslash \pc{u}$}
        \STATE{Consider file index $\set{W} \leftarrow \set{M}\backslash\pc{t}$}
        \STATE{$D_{\set{F},u}^k \leftarrow D_{\set{F},u}^k \xor I_{\set{W},u}^t$ }
        \ENDFOR
      \ENDFOR
    \STATE{$I_{\set{F}}^k \leftarrow$ Merge $\pc{ D_{\set{F},u}^k : u \in \set{F} }$}
    \STATE{Store $I_{\set{F}}^k$}
    \ENDFOR
  \end{algorithmic}
  \caption{Decoding}
  \label{alg:decode}
\end{algorithm}

\subsection{Reduce}
After the Decoding stage, we note that Node $k$ has obtained all KV pairs in the partition $P_k$, for all $k \in \set{K}$. In particular, the KV pairs $\pc{I_{\cal S}^k :k \in \set{S}, |\set{S}|=r}$ are obtained locally in the Map stage, and the KV pairs $\pc{I_{\cal S}^k :k \notin \set{S}, |\set{S}|=r}$ are obtained in the above Decoding stage. 

In this final stage, Node $k$, $k=1,\ldots,K$, performs the Reduce process as described in Section \ref{sec:reduce} for the \texttt{TeraSort} algorithm, sorting the KV pairs in partition $P_k$ locally.  



\section{Evaluation}
\label{sec:evaluation}
We imperially demonstrate the performance gain of \texttt{CodedTeraSort} through experiments on Amazon EC2 clusters.  In this section, we first present the choices we have made for the implementation. Then, we describe experiment setup. Finally, we discuss the experiment results.

\subsection{Implementation Choices}

We first describe the following implementation choices that we have made for both \texttt{TeraSort} and \texttt{CodedTeraSort} algorithms.

\emph{Data Format:} All input KV pairs are generated from \texttt{TeraGen}~\cite{TSpackage} in the standard Hadoop package.  Each input KV pair consists of a $10$-byte key and a $90$-byte value. A key is a $10$-byte unsigned integer, and the value is an arbitrary string of $90$ bytes. The KV pairs are sorted based on their keys, using the standard integer ordering.

\emph{Platform and Library:} We choose Amazon EC2 as the evaluation platform.  We implement both \texttt{TeraSort} and \texttt{CodedTeraSort} algorithms in \texttt{C++}, and use Open MPI library \cite{openMPI} for communications among EC2 instances.

\begin{figure}
  \centering
  \includegraphics[scale=1.0]{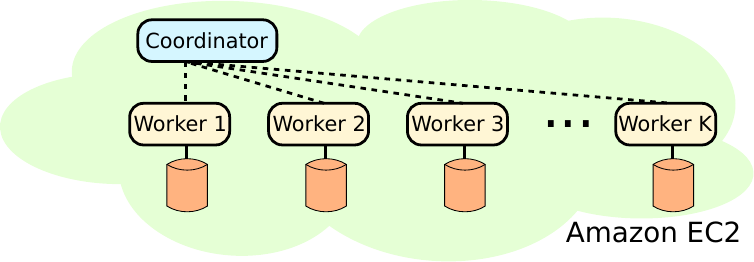}
  \caption{The coordinator-worker system architecture.}
  \label{fig:system}
\end{figure}

\emph{System Architecture:} As shown in Fig. \ref{fig:system}, we employ a system architecture that consists of a coordinator node and $K$ worker nodes, for some $K \in \mathbb{N}$. Each node is run as an EC2 instance. The coordinator node is responsible for creating the key partitions and placing the input files on the local disks of the worker nodes. The worker nodes are responsible for distributedly executing the stages of the sorting algorithms. 

\emph{In-Memory Processing:} After the KV pairs are loaded from the local files into the workers' memories, all intermediate data that are used for encoding, decoding and local sorting are persisted in the memories, and hence there is no disk I/O involved during the executions of the algorithms.

\begin{figure}
  \centering
  \includegraphics[scale=0.65]{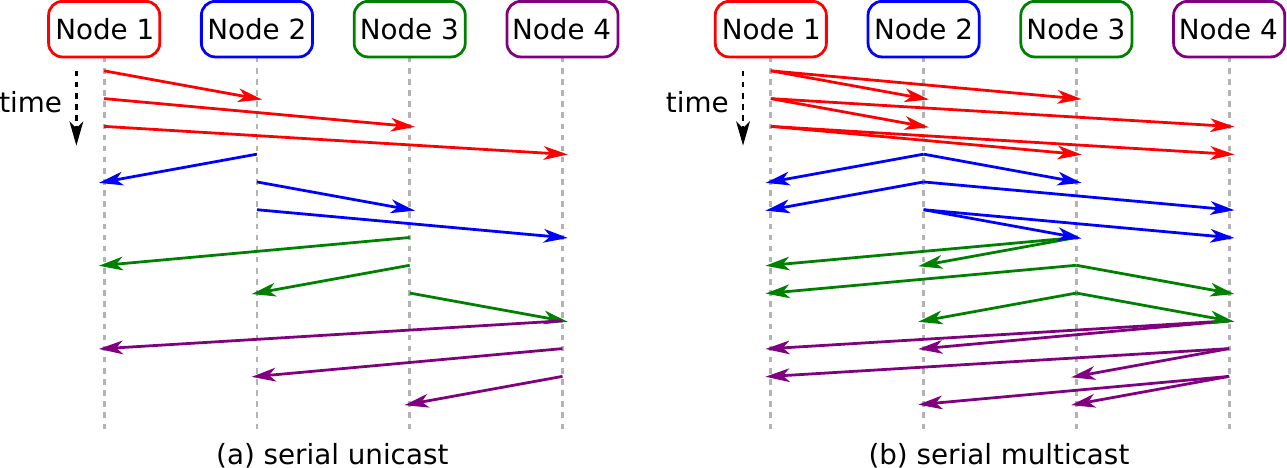}
  \caption{(a) Serial unicast in the Shuffle stage of \texttt{TeraSort}; a solid arrow represents a unicast. (b) Serial multicast in the Multicast Shuffle stage of \texttt{CodedTeraSort}; a group of solid arrows starting at the same node represents a multicast.}
  \label{fig:serial}
\end{figure}

In the \texttt{TeraSort} implementation, each node sequentially steps through Map, Pack, Shuffle, Unpack, and Reduce stages. The Map, Shuffle, and Reduce stages follow the descriptions in Section \ref{sec:TeraSort}. In the Reduce stage, the standard sort \texttt{std::sort} is used to sort each partition locally. To better interpret the experiment results, we add the Pack and the Unpack stages to separate the time of serialization and deserialization from the other stages.  The Pack stage serializes each intermediate value to a continuous memory array to ensure that a single TCP flow is created for each intermediate value (which may contain multiple KV pairs) when \texttt{MPI\_Send} is called\footnote{Creating a TCP flow per KV pair leads to inefficiency from overhead and convergence issue.}.  The Unpack stage deserializes the received data to a list of KV pairs.  In the Shuffle stage, intermediate values are unicast serially, meaning that there is only one sender node and one receiver node at any time instance.  Specifically, as illustrated in Fig. \ref{fig:serial}(a), Node $1$ starts to unicast to Nodes 2, 3, and 4  back-to-back.  After Node $1$ finishes, Node $2$ unicasts back-to-back to Nodes 1, 3, and 4.  This continues until Node 4 finishes. 

In the \texttt{CodedTeraSort} implementation, each node sequentially steps through CodeGen, Map, Encode, Multicast Shuffling, Decode, and Reduce stages.  The Map, Encode, Multicast Shuffling, Decode, and Reduce stages follow the descriptions in Section \ref{sec:codedTeraSort}.  In the CodeGen (or code generation) stage, firstly, each node generates all file indices, as subsets of $r$ nodes. Then each node uses \texttt{MPI\_Comm\_split} to initialize $\binom{K}{r+1}$ multicast groups each containing $r+1$ nodes on Open~MPI, such that multicast communications will be performed within each of these groups.  The serialization and deserialization are implemented respectively in the Encode and the Decode stages.  In Multicast Shuffling, $\texttt{MPI\_Bcast}$ is called to multicast a coded packet in a serial manner, so only one node multicasts one of its encoded packets at any time instance.  Specifically, as illustrated in Fig. \ref{fig:serial}(b), Node 1 multicasts to the other 2 nodes in each multicast group Node 1 is in. For example, Node 1 first multicasts to Node 2 and 3 in the multicast group $\{1,2,3\}$.  After Node $1$ finishes, Node $2$ starts multicasting in the same manner. This process continues until Node $4$ finishes.

\subsection{Experiment Setup}
We conduct experiments using the following configurations to evaluate the performance of \texttt{CodedTeraSort} and \texttt{TeraSort} on Amazon EC2:
\begin{itemize}
    \item The coordinator runs on a r3.large instance with $2$ processors, $15$ GB memory, and $32$ GB SSD.
    \item Each worker node runs on an m3.large instance with $2$ processors, $7.5$ GB memory, and $32$ GB SSD.
    \item The incoming and outgoing traffic rates of each instance are limited to $100$ Mbps.\footnote{This is to alleviate the effects of the bursty behaviors of the transmission rates in the beginning of some TCP sessions. The rates are limited by traffic control command \texttt{tc} \cite{TCcommand}.}
    \item $12$ GB of input data (equivalently $120$ M KV pairs) is sorted.
\end{itemize}

We evaluate the run-time performance of \texttt{TeraSort} and \texttt{CodedTeraSort}, for different combinations of the number of workers $K$ and the parameter $r$. All experiments are repeated $5$ times, and the average values are reported.




\subsection{Experiment Results}

\begin{table*}[!t]
\centering
\caption{Sorting $12$ GB data with $K = 16$ worker nodes and 100 Mbps network speed}
  \label{tlb:K16}
  \begin{tabular}{|c|c|c|c|c|c|c|c|c|}
    \hline
             & CodeGen & Map & Pack/Encode & Shuffle & Unpack/Decode & Reduce & Total Time & Speedup \\
             & (sec.)   & (sec.) & (sec.) & (sec.) & (sec.) & (sec.) & (sec.) & \\\hline
    \texttt{TeraSort}:           &  --   & 1.86  & 2.35 & 945.72  & 0.85 & 10.47 & 961.25 & \\
    \texttt{CodedTeraSort}: $r=3$& 6.06	 & 6.03  & 5.79 & 412.22  & 2.41 & 13.05 & 445.56 & 2.16$\times$ \\
    \texttt{CodedTeraSort}: $r=5$& 23.47 & 10.84 & 8.10 & 222.83  & 3.69 & 14.40 & 283.33 & 3.39$\times$ \\\hline
  \end{tabular}
\end{table*}

\begin{table*}[!t]
\centering
\caption{Sorting $12$ GB data with $K = 20$ worker nodes and 100 Mbps network speed}
  \label{tlb:K20}
  \begin{tabular}{|c|c|c|c|c|c|c|c|c|}
    \hline
             & CodeGen & Map & Pack/Encode & Shuffle & Unpack/Decode & Reduce & Total Time & Speedup \\
             & (sec.)   & (sec.) & (sec.) & (sec.) & (sec.) & (sec.) & (sec.) & \\\hline
    \texttt{TeraSort}:           &  --  & 1.47 & 2.00 & 960.07 & 0.62 & 8.29 & 972.45 & \\
    \texttt{CodedTeraSort}: $r=3$& 19.32 & 4.68 & 4.89 & 453.37 & 1.87 & 9.73 & 493.86 & 1.97$\times$ \\
    \texttt{CodedTeraSort}: $r=5$& 140.91 & 8.59 & 7.51 & 269.42 & 3.70 & 10.97 & 441.10 & 2.20$\times$ \\\hline
  \end{tabular}
\end{table*}

The breakdowns of the execution times with $K=16$ workers and $K=20$ workers are shown in Tables \ref{tlb:K16} and \ref{tlb:K20} respectively.  We observe an overall $1.97\times$-$3.39\times$ speedup of \texttt{CodedTeraSort} as compared with \texttt{TeraSort}.  From the experiment results we make the following observations:
\begin{itemize}
    \item For \texttt{CodedTeraSort}, the time spent in the CodeGen stage is proportional to 
    $\binom{K}{r+1}$, which is the number of multicast groups.  
    \item The Map time of \texttt{CodedTeraSort} is approximately $r$ times higher than that of \texttt{TeraSort}. This is because that each node hashes $r$ times more KV pairs than that in \texttt{TeraSort}.  Specifically, the ratios of the \texttt{CodedTeraSort}'s Map time to the \texttt{TeraSort}'s Map time from Table \ref{tlb:K16} are $6.03/1.86 \approx 3.2$ and $10.84/1.86 \approx 5.8$, and from Table \ref{tlb:K20} are $4.68/1.47 \approx 3.2$ and $8.59/1.47 \approx 5.8$.
    \item While \texttt{CodedTeraSort} theoretically promises a factor of more than $r \times$ reduction in shuffling time, the actual gains observed in the experiments are slightly less than $r$. For example, for an experiment with $K=16$ nodes and $r=3$, as shown in Table \ref{tlb:K16}, the speedup of the Shuffle stage is $945.72/412.22 \approx 2.3 < 3$. This phenomenon is caused by the following two factors. 1) Open MPI's multicast API ($\texttt{MPI\_Bcast}$) has an inherent overhead per a multicast group, for instance, a multicast tree is constructed before multicasting to a set of nodes.  2) Using the $\texttt{MPI\_Bcast}$ API, the time of multicasting a packet to $r$ nodes is higher than that of unicasting the same packet to a single node. In fact,
    as measured in~\cite{Lee2015Bcast}, the multicasting time increases logarithmically with $r$.

    \item The sorting times in the Reduce stage of both algorithms depend on the available memories of the nodes.  \texttt{CodedTeraSort} inherently has a higher memory overhead, e.g., it requires persisting more intermediate values in the memories than \texttt{TeraSort} for coding purposes, hence its local sorting process takes slightly longer.  This can be observed from the Reduce column in Tables \ref{tlb:K16} and \ref{tlb:K20}.
    \item The total execution time of \texttt{CodedTeraSort} improves over \texttt{TeraSort} whose communication time in the Shuffle stage dominates the computation times of the other stages.
\end{itemize}

Further, we observe the following trends from both tables:

\emph{The impact of redundancy parameter $r$:} As $r$ increases, the shuffling time reduces substantially by approximately $r$ times. However, the  Map execution time increases linearly with $r$, and more importantly the CodeGen time increases as $\binom{K}{r+1}$. Hence, for small values of $r$ ($r < 6$) we observe overall reduction in execution time, and the speedup increases. However, as we further increase $r$, the CodeGen time will dominate the execution time, and the speedup decreases. Hence, in our evaluations, we have limited $r$ to be at most $5$.\footnote{The redundancy parameter $r$ is also limited by the total storage available at the nodes. Since for a choice of redundancy parameter $r$, each piece of input KV pairs should be stored at $r$ nodes, we can not increase $r$ beyond $\frac{\text{total available storage at the worker nodes}}{\text{input size}}.$}

\emph{The impact of worker number $K$:} As $K$ increases, the speedup decreases. This is due to the following two reasons. 1) The number of multicast groups, i.e., $\binom{K}{r+1}$, grows exponentially with $K$, resulting in a longer execution time of the CodeGen process. 2) When more nodes participate in the computation, for a fixed $r$, less amount of KV pairs are hashed at each node locally in the Map stage, resulting in less locally available intermediate values and a higher communication load.   
    
In addition to the results in Tables \ref{tlb:K16} and \ref{tlb:K20}, we have performed more experiments, and listed their results on~\cite{CTSweb}. From those results, we observe that \texttt{CodedTeraSort} achieved up to $4.11\times$ speedup.

\section{Conclusion and Future Directions}

In this paper, we integrate the principle of a recently proposed Coded MapReduce scheme into the \texttt{TeraSort} algorithm, developing a novel distributed sorting algorithm \texttt{CodedTeraSort}. \texttt{CodedTeraSort} specifies a structured redundant placement of the input files that are to be sorted, such that the same file is repetitively processed at multiple nodes. The results of this redundant processing enable in-network coding opportunities that substantially reduce the load of data shuffling. We also empirically demonstrate the significant performance gain of \texttt{CodedTeraSort} over \texttt{TeraSort}, whose execution is limited by data shuffling.

Finally, we highlight three future directions of this work.
\begin{itemize}
    \item \emph{Beyond Sorting Algorithms.} Having successfully demonstrated the impact of coding in improving the performance of  \texttt{TeraSort}, we can apply the coding concept to develop coded versions of many other distributed computing applications whose performance is limited by data shuffling (e.g., Grep, SelfJoin). In particular, mobile machine learning applications like mobile augmented reality and recommender systems are of special interest since the communications through wireless links are much slower. In recent works, the impact of coded distributed computing on wireless distributed computing has been theoretically studied in~\cite{edgeComp1,edgeComp2}.
\item \emph{Scalable Coding.} We observe from the experiment results
that the coding complexity (i.e., the time spent at CodeGen stage) increases as $\binom{K}{r+1}$. Hence, as the redundancy parameter $r$ gets large  the coding overhead (including the time spent in generating the coding plan, encoding, and decoding) becomes comparable with or even longer than the time spent in Map and Reduce stages. It is of great interest to design  efficient and scalable coding procedures to maintain a low coding overhead.
\item \emph{Asynchronous Execution.} In the experiments, we executed the stages of the computation one after another in a synchronous manner. Also, the data shuffling was performed serially such that only one node is communicating (unicasting for \texttt{TeraSort} and multicasting for \texttt{CodedTeraSort}) at a time. It is interesting to explore the impact of coding in an asynchronous setting with parallel communications.  
\end{itemize}


\bibliographystyle{ieeetr}
\bibliography{main.bib}

\end{document}